\newcommand{\be}{\begin{equation}}
\newcommand{\bea}{\begin{eqnarray}}
\newcommand{\ee}{\end{equation}}
\newcommand{\eea}{\end{eqnarray}}
\newcommand{\qe}{\varepsilon}

\newcommand{\qs}{\sigma}

\newcommand{\tr}{{\rm tr}\;}

\newcommand{\dagg}{^{\dag}}
\newcommand{\prt}{\partial}

\newcommand{\fr}[2]{{\textstyle \frac{#1}{#2}}}

\newcommand{\scez}{\setcounter{equation}{0}}
\newcommand{\ns}{\scez\section}

\documentclass[preprint,aps,eqsecnum,prb,tighten]{article}
\usepackage[dvips]{graphicx}

\begin{document}

%\draft

\title{ The large N theory exactly reveals \\ the quantum Hall effect and 
$\theta$ renormalization}

\author{A.M.M. Pruisken, M.A. Baranov and M. Voropaev}
\maketitle
\vspace{-1cm}
 
\begin{center}
{\it Institute for Theoretical Physics, University of Amsterdam,
Valckenierstraat 65, \\ 1018 XE Amsterdam, The Netherlands}
\end{center}

%\maketitle

\vspace{1cm}

\begin{abstract}
\noindent
We revisit the $\theta$ dependence in $CP^{N-1}$ model with large $N$. We
study the consequences of a recently discovered new ingredient of the 
instanton vacuum, i.e. the massless chiral edge excitations.
Contrary to the previous believes, our results demonstrate that the large $N$ expansion
displays all the fundamental features of the quantum Hall effect. This 
includes the {\em robust} quantization of the Hall conductance, as well as 
the appearance of massless bulk excitations at $\theta = \pi$. We conclude that
the quantum Hall effect is a generic feature of the instanton vacuum, i.e. for
any number of field components and not just for the theory in the replica limit alone.
\end{abstract}

PACS numbers 72.10.-d, 73.20.Dx, 73.40.Hm

\vfill

\pagebreak[4]

\tableofcontents

\newpage 

\ns{Introduction}

\subsection{The large $N$ theory}
 
The $CP^{N-1}$ model has historically been of interest  because of its similarities
with $QCD$. Like $QCD$, the $CP^{N-1}$ theory is scale invariant, 
asymptotically free, it possesses instantons as well as a $\theta$ parameter.
Unlike $QCD$, however, it has a tractable large $N$ expansion that can be used to explore and 
investigate the important non perturbative aspects of the theory.

The large $N$ theory in 1+1 space time dimension displays, as is well known,  
impressive features such as mass generation and infrared slavery. 
It is also one of the very few places where detailed studies can be made of the otherwise
unknown $\theta$ dependence in scale invariant theories. The results seem to
indicate that the theory has always a massgap, for all values of $\theta$. On the otherhand, 
the vacuum free energy develops a "gusp" (a first order phase transition) as $\theta$ passes
through odd multiples of $\pi$. 

The results of the large $N$ theory have historically set the stage for elaborate discussions on 
such topics like the meaning of discrete topological sectors in scale invariant theories,
the role of instantons etc (Ref [1]-[4]).  

\subsection{The quantum Hall effect}
It is important to keep in mind, however, that much of the large $N$ theory, as it was studied
a long time ago, is in direct conflict with one of the most
interesting applications of the {\em instanton vacuum} concept in Condensed Matter Physics,
namely the quantum Hall effect ({\em qHe}). In contrast to QCD and other theories where the 
exact meaning of the $\theta$ parameter has remained 
somewhat obscure and experiments are impossible, the situation of the {\em qHe} is much better.
In this case, $\theta$ is associated with a physical observable, namely the Hall conductance. 
What emerges from the experimental phenomenon of the {\em qHe} is, in fact, a very different and 
previously unrecognized behavior of the strong coupling phase. 
In particular, {\em massless} excitations must exist 
at $\theta = \pi$, and the theory must display a {\em quantum Hall
phase} with a massgap otherwise. Here, the phrase {\em quantum Hall phase} means that 
{\em robust topological quantum numbers} exist that explain the stability and
precision of the {\em qHe}.
These topological quantum numbers, for all we know, can be identified as the {\em renormalized}
value of the instanton angle $\theta$ (Ref [5]-[9]).

\subsection{$\theta$ renormalization}
The concept of $\theta$ {\em renormalization} is, in many ways,
the most profound and novel idea that arose from the instanton vacuum approach to the {\em qHe}.
On the one hand,
it explains in a natural fashion how von Klitzing's discovery can be reconciled with the perturbative quantum
theory of conductances (Anderson localization). On the other hand, it is the standard
Statistical Mechanics way of describing the topological singularities of the theory at $\theta = \pi$. 
Most of our present understanding of the {\em qHe} is based on
these two distinctly different physical aspects of $\theta$ renormalization.

\subsection{Gapless edge modes}
In a recent series of papers it was shown that a fundamental ingredient of the theory has
previously been overlooked. It has turned out that the instanton vacuum generally displays a
spectrum of {\em gapless edge excitations}. These edge excitations are identical to the
massless chiral edge modes that are well known to exist in quantum Hall systems.
This important aspect of the
problem has been studied in detail and exploited in Ref. 14, 15  for the purpose of
deriving the complete Luttinger
liquid theory for edge excitations in abelian quantum Hall states, from first principles.

The problem of massless edge modes generally complicates the phenomenon of
mass generation in asymptotically free field theories. It means that the instanton vacuum
{\em dynamically} generates distinctly different spectra for bulk and edge excitations and
this, in turn, has important topological consequences. 
In this paper we shall elaborate further on this
point and show how it affects the fundamental subjects of {\em renormalization}, 
the {\em Kubo formalism} for the conductances, the {\em background field methodology} as well as such  
problems like the {\em quantization} of topological charge. (Ref. 11, 17)

\subsection{Gapless bulk modes}

The existence of massless edge excitations also affects the properties of the bulk excitations at
$\theta = \pi$. For example, in Ref. 14 we addressed the longstanding problem of long ranged potential
fluctuations in realistic quantum Hall devices which act as an inexhaustible source for extended edge states
that percolate throughout the bulk of the system. The instanton vacuum approach to the {\em qHe} 
can be used to show that the crossing over between semi classical ideas (percolation) and
Anderson localization involves, generally speaking, arbitrary 
small temperatures.

It is natural to expect that the first order singularity, as
found in the $\theta$
dependence of  the $CP^{N-1}$ model with large $N$, is unstable with
respect to th
formation of massless edge modes in the bulk of the system. One can
stretch the analogy 
with percolation a little further and conclude that at the system, at
$\theta = \pi$, is likely
to {\em delocalize} due to the presence of edge excitations which always
appear at the
interface of two different quantum Hall phases. This mechanism of 
generating massless
excitations at $\theta = \pi$ was previously not recognized, and until to
date, the conflict
between the large $N$ theory and the {\em qHe} has remained largely
obscure.

This motivates us to revisit the
large $N$ expansion and use it as a possible example where the
non-perturbative features 
of quantum Hall physics can be explored and investigated in detail.

The main conclusion of this paper is  that the large $N$
theory displays,
indeed, a diverging correlation length or a gapless phase at $\theta =
\pi$ . Along
with that, we show that the large $N$ expansion is a unique laboratory
for studying
the quantum Hall plateau transitions. Exact scaling functions are
obtained, and the
corrections to quantization are found to be exponentially small in the
{\em
area}of the system, rather than the linear dimension.

\subsection{Outline of this paper}

The physics of the {\em qHe} is intimidly related to such 
longstanding problems like {\em quantization} of topological charge, the role of {\em instantons} in scale invariant
theories etc.  Many of these problems have been interpreted incorrectly
in the large $N$ analyses before. For this reason we shall start out, in the first part of this paper (Sections 2 and 3), 
with a brief review of the general theory and we embark on the subtle effects of massless edge excitations. 

As is well known, the general theory is defined on a generalized
$CP^{N-1}$
manifold, the $\frac{U(N_1 +N_2 )}{U(N_1 ) \times U(N_2 )}$ non linear
sigma model, and the theory of the {\em qHe} is formally obtained by
putting
the number of field components equal to zero, $N_1 = N_2 =0$ ({\em
replica limit}).

The general theory shares all the aforementioned features with $QCD$, and
this includes
the most interesting case where $N_1 =N_2 =0$. Fig. 1 indicates how the
nature of
the topological singularities at $\theta = \pi$ depend on the value of
$N_1$ and $N_2$.
We return to this point at the end of Section 3 where we review the
results obtained from instanton calculus.

Since massless edge excitations actually appear as a non-perturbative phenomenon in the strong coupling regime, 
it is generally not quite obvious how one can
distinguish between the bulk and the edge degrees of freedom in the action.
This will be discussed of Section 3 where we introduce distinctly different actions for bulk and edges, by making use 
of the microscopic origins of the instanton vacuum approach to the {\em qHe}. 

The physics of the electron gas turns out to be particularly helpful in establishing the
topological significance of bulk and edge quantities. For example, it provides a natural way of splitting 
the non-linear $\sigma$ model field 
variables ($Q$) in {\em bulk} components which generally have a {\em quantized} topological
charge, and {\em edge} components which always have an {\em unquantized} topological charge.
Moreover, it motives the idea of formulating an effective action for the (massless, chiral) edge modes, 
that is obtained by formally eliminating the bulk 
degrees of freedom, which carry a quantized topological charge, from the theory. 

\begin{figure}
\begin{center}
\includegraphics[height=4cm]{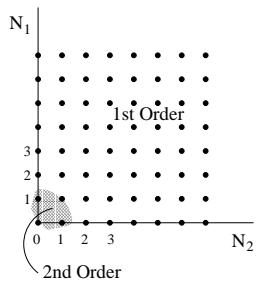}
\end{center}
\caption{Nature of the transition at $\theta = \pi$}
\end{figure}

At this stage, several
aspects of the theory become simultaneously important. We mention in particular, the concept of 
mass generation (in the bulk of the system), the background field
methodology which has been introduced and analyzed in great detail for renormalization group purposes in our earlier work,
as well as the Kubo formalism for the conductances. They all are an integral part of one of the main statements
made in this paper, which says that {\em quantum Hall phases} are, in fact, a {\em universal} strong coupling feature of the
topological concept of an {\em instanton vacuum}.

Armed with the insights gained in Section 3, we next specialize to the $\theta$ dependence of the $CP^{N-1}$ theory
with large $N$ in Section 4. We follow up on the usual formulation which is expressed in terms of $U(1)$ gauge fields. 
Elementary considerations indicate that the aforementioned separation between edge and bulk physics directly
displays, by construction, Coleman's heuristic ideas for having periodicity in the parameter $\theta$.

At the same time we show the {\em qHe} is generated by the effective action of the edge modes, and how
a massless phase or a diverging length scale appears at $\theta = \pi$.

For the remainder of this paper (Section 5), we study a more systematic approach which
leads to a representation of the problem in terms of the 1D Coulomb gas with statistical
charges at opposite ends of the universe. This representation is used to extract detailed information
on the renormalization of the theory which now includes the entire regime from weak to 
strong coupling. 

The Coulomb gas results provide valuable insight
into the role of instantons in scale invariant theories. 
Although instantons are usually assumed to be irrelevant at large $N$, 
it is nevertheless possible to cast the results for the free energy and 
the renormalization group $\beta$ functions into a {\em quasi instanton}
form. We discuss these results and show that they accurately describe the principle 
features of an exact theory, including the transition at $\theta = \pi$.

The idea of {\em quasi instantons} is in some ways similar to the idea of {\em quasi particles} in 
Fermi liquid theory. It simply says that inspite of the strong interactions between the 
semi classical topological objects in the strong coupling regime, it is nevertheless 
possible to describe the many body system in terms of a gas of nearly free quasi instantons. 
Just like in Fermi liquid theory, there is generally no simple relation between the
semi classical instantons and the quasi objects.

We end this paper with a discussion in Section 6.

\section{Non-linear $\sigma$ model}
\subsection{Introduction}
Recall the effective replica field theory  or non-linear $\sigma$ model
for completely disordered 2D electrons in a perpendicular
magnetic field $B$. It involves the matrix field variables $Q$ defined on the symmetric space
$\frac{U(N_1 +N_2 )}{U(N_1 ) \times U(N_2 )}$. The complete action can be written as

\be
 	S_{eff} [Q] =  -\fr{1}{8}\qs^0_{xx}\int d^2 x \tr \prt_\mu Q \prt_\mu Q 
 	+\fr{1}{8}\qs^0_{xy} \int d^2 x \tr \qe_{\mu\nu}Q\prt_\mu Q\prt_\nu Q  + \rho \omega \int d^2 x tr\Lambda Q .
\ee
Here, the dimensionless parameters $\sigma_{xx}^0$, $\sigma_{xy}^0$ and $\rho$ represent the {\em 
meanfield theory} for
longitudinal conductance, the Hall conductance and the density of states respectivily, whereas $\omega$
stands for the external frequency. 

The unitary matrices $Q$ obey the non linear constraint $Q^2 =1$. They are represented by

\be
Q= {T\dagg} \Lambda T 
\ee
where $T\epsilon U(N_1 + N_2 )$ and $\Lambda$ is a diagonal matrix with $N_1$ elements $+1$ 
and $N_2$ elements $-1$. 

The topological charge $q(Q)$ of the theory can now be written in two equivalent ways

\be
q (Q) = \frac{1}{16\pi i} \int d^2 x \tr \qe_{\mu\nu}Q\prt_\mu Q \prt_\nu Q = 
\frac{1}{2\pi i} \oint dx tr T\prt_x {T\dagg} \Lambda ,
\ee
where the surface integral is over the edge of the disordered sample.

\subsection{Chiral edgemodes}
The theory simplifies dramatically in the limit of very strong $B$ where the Landau bands are separated by
an energy gap. If we specify the Fermi 
level to lie in a Landau gap than both $\sigma_{xx}^0$ and $\rho$ vanish and $\sigma_{xy}^0 = m$
is an integer, equal to the number of completely filled Landau levels. 
The effective action then reduces to a 1D edge theory which can be written as follows

\be
 	S_{edge} [Q] = \fr{m}{2} \oint d x \tr T \prt_x {T\dagg} \Lambda + \rho_{edge} 
 	\omega \oint d x tr\Lambda Q .
\ee
Here, the term with $\rho_{edge}$ indicates that although there are no levels in the bulk, there still is a
finite density of chiral edge states at the Fermi energy that carry the Hall current.
In order to see which part of the $T$ matrix fields in $S_{edge}$ can be generally taken as an edge degree of freedom
we split the 2D matrix field variables $Q$ into ("bulk") 
components $Q_0$ and ("edge") components $t$ according to

\be
Q = {t\dagg} Q_0 t .
\ee
Here, the $Q_0$ are defined with {\em spherical boundary conditions} $Q_0 |_{edge} =\Lambda$. The unitary rotations $t$
then generally represent the {\em fluctuations} about the spherical boundary conditions.

It is easy to see that $S_{eff}$ depends on the $t$ matrix field variables only. Explicitly written out
we have

\be
 	S_{edge} [{t\dagg} Q_0 t] = 2\pi i m q(Q_0 )+ \fr{m}{2} \oint d x \tr t \prt_x {t\dagg} \Lambda 
 	+ \rho_{edge} 
 	\omega \oint d x tr\Lambda ({t\dagg} \Lambda t) .
\ee
This expression contains the topological charge $q(Q_0 )$ which is, by construction, integer valued. It gives 
rise to phase factors that are immaterial, but physically it stands for an integer number of edge electrons that
have crossed the Fermi level.

In Ref. 14, 15 is was shown that this 1D theory $S_{edge}$ is exactly solvable and completely equivalent to the theory
of massless chiral edge bosons. Some important correlations are given by 

\be
<Q> |_{edge} = \Lambda
\ee
and

\be
<Q_{+-}^{\alpha\beta} (x) Q_{-+}^{\beta\alpha} (x') > |_{edge} =\frac{4}{m} \Theta(x'-x) e^{-\frac{4\rho_{edge} \omega}{m}
(x'-x)}
\ee
Notice that these edge correlations are the same, 
independent of the number of field components $N_1$ and $N_2$ in the theory.
This feature was exploited and extended in Ref. 14, 15 for the purpose of deriving the complete Luttinger liquid
theory of the abelian quantum Hall edge states, from first principles.

\section{Disentangling bulk and edge modes}
Notice that on
the basis of asymptotic freedom alone one would expect that the general theory (Eq. 1) generates a massgap.
The results of the previous Section seem to indicate, however, that massless excitations do exist in
the regime of strong coupling and they are confined to the edge of the instanton vacuum. It is now natural
to suppose that the {\em qHe} is a general consequence of two distinctly different strong coupling aspects
of the theory, namely the bulk excitations ($Q_0$) which generate a mass in the limit of large distances 
on the one hand, and the edge excitations ($t$) which remain massless on the other.
  
In order to generaly distinguish between the bulk and edge pieces
of the problem, it is important to get some help from the microscopic origins of the action. In what
follows we shall make use of some of the details of the derivation of the theory.

First, it is convenient
(although not strictly necessary) to work in the limit of strong Landau quantization (strong $B$), since this
fascilitates a discussion of important aspects like particle hole symmetry. Moreover, the expression for 
the parameter $\sigma_{xy}^0$ simplifies in this case and it is equal to the filling fraction $\nu$ of
the Landau bands. The quantity $\nu$ is an adjustible parameter in this problem and it is convenient to think of
all physical quantities as being a function of $\nu$. 

Secondly, in order to see that the bulk and edge pieces in the problem are physically distinguishable, it is
helpful to consider a system where the confining edge is "spatially seperated" from a homogenously disordered 
interior. This kind of thought experiment has been frequently used in Ref. 14, 15 and here it means that the action 
$S_{eff}$ can written as the sum of a bulk part $S_b$ and an edge part $S_e$ as follows

\be
	S_{eff} [Q] = S_b [Q] + S_e [Q] ,
\ee
where
\be
 	S_b [Q] = -\fr{1}{8}\qs^0_{xx}\int d^2 x \tr \prt_i Q \prt_i Q 
 	+\fr{\theta(\nu)}{16\pi} \int d^2 x \tr \qe_{ij}Q\prt_i Q\prt_j Q + \rho \omega \int d^2 x tr\Lambda Q
\ee
and where $S_e$ is the typical result for an isolated confining edge (Eq. 4)
\be
 	S_e [Q] = \fr{m(\nu)}{2} \oint d x \tr T \prt_x {T\dagg} \Lambda + \rho_{edge} 
 	\omega \oint d x tr\Lambda Q .
\ee
\begin{figure}
\begin{center}
\includegraphics[height=4cm]{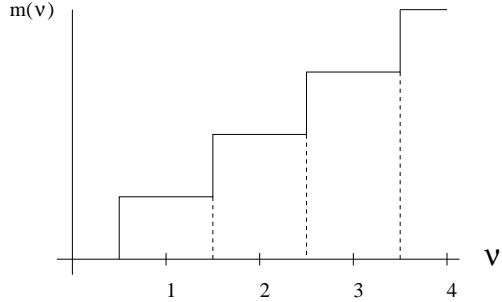}
\end{center}
\caption{The edge part of $\sigma_{xy}^0$ with varying $\nu$ (see text)}
\end{figure}
The most important physical quantities are $m(\nu)$ and $\theta(\nu)$ which are defined by
\be
	m(\nu) = \sum_{n=0} \Theta (\nu -\fr{2n+1}{2} ) ; \theta (\nu) = 2\pi (\nu - m(\nu) ) .
\ee
Notice that $m(\nu) + \frac{\theta(\nu)}{2\pi} = \nu = \sigma_{xy}^0$ such that the sum of $S_b$ and $S_e$ 
precisely equals the original action, Eq. 1, except for the quantity $\rho_{edge}$ which indicates that the
chiral edge levels are distincly different from the two dimensional bulk levels.  
The edge quantity $m(\nu)$ is related to the one dimensional density of edge states $\rho_{edge}$ according to

\be
m(\nu) = 2v_d \cdot \rho_{edge}
\ee
where $v_d$ stands for the drift velocity of the chiral edge electrons.

\begin{figure}
\begin{center}
\includegraphics[height=4cm]{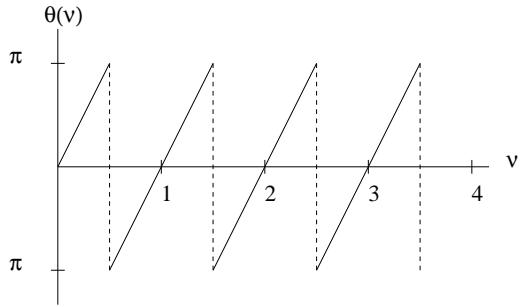}
\end{center}
\caption{The bulk part of $\sigma_{xy}^0$ with varying $\nu$ (see text).}
\end{figure}

\subsection{Particle hole symmetry}
One way to demonstrate that $S_b$ is indeed the action for the bulk is by considering {\em particle hole symmetry} in the
separate Landau bands(Ref. 12). By particle hole symmetry we mean the simple relation that exists between having
an {\em empty} Landau band that is partially filled with electrons and a {\em full} Landau band that is
partly filled with holes.  
For example, for a partly filled Landau level with index $n$, the number of particles and the number of holes 
are interchanged if one interchanges the
filling fractions $\nu_+ = n+\frac{1}{2} +\epsilon$ and $\nu_- = n+\frac{1}{2} -\epsilon$ respectivily 
($|\epsilon| \le \frac{1}{2}$). From the mean field results in strong $B$ we have for the density of bulk states

\be
\rho (\nu_+) = \rho (\nu_-) 
\ee
as expected. On the other hand both $\sigma_{xx}^0$ and $\theta (\nu)$ are non zero only if $\rho$ is non zero 
whereas (see also Fig. 3)

\be
\sigma_{xx}^0 (\nu_+) = \sigma_{xx}^0 (\nu_-) ,
\theta (\nu_+) = -\theta (\nu_-) 
\ee
indicating that we are dealing with bulk quantities. 

Finally, it is important to remark that the contruction of $S_b$ and $S_e$ is not unique to the physics of strong $B$
and a formal distinction between edge and bulk actions can be made quite
generally. However, the significance of introducing of discontinuous funtions like $m(\nu)$ and $\theta(\nu)$ 
is not quite obvious in general and the physics of strong $B$ serves, in many ways, as an important guide.  
Moreover, we shall show, in the next Section, that the symmetries of the strong $B$ problem are, in fact, 
retained by the renormalization of the theory. The results set the stage for a strong coupling scenario
that shall be further explored for the $CP^{N-1}$ theory with large $N$.

\subsection{Background field methodology}
%It is important to stress that the contributions of the sample edge, $S_e$ or $m(\nu)$, have
%quite generally been overlooked, not only in the context of the {\em qHe}, but also in the 
%previous work on the $CP^{N-1}$ theory with large $N$ which is the main subject of the remainder of
%this paper. It has, however, fundamental consequences for the theory that does  
%not make much physical sense otherwise. 

Let us proceed and introduce the split in bulk and edge
degrees of freedom (Eq. 5). The main idea next is to formulate an effective action for the
field variables $t$ that is obtained by performing the integral over the bulk modes $Q_0$.
Specifically

\be
	e^{S_{eff} [t]} = e^{S_e [t]} \int D[Q_0 ] e^{S_b ({t\dagg} Q_0 t )} .
\ee
>From now onward, we put the external frequencies $\omega$ equal to zero and work with a finite system size
instead. It is easy to understand what is generally meant by this construction and how it generates the
{\em qHe}. Notice first that $S_b$ vanishes identically in the simple case where the Fermi level is located
in a Landau gap (Section 2.2). $S_{eff}$ has then the same meaning as $S_e$ with $m(\nu=1,2,3, ...)$ representing the
Hall conductance.

Next, by moving slightly away from precisely integer filling fractions $\nu$,
all parameters $\sigma_{xx}^0$, $\theta(\nu)$ and $\rho$ in $S_b$ gain small values different from zero.
The problem of evaluating $S_{eff}$ now suddenly becomes a very complex strong coupling problem that can hardly
be treated analytically. 

If, however, it is reasonable to expect that the theory $S_b$ with a small value of
$\theta(\nu)$ generates a massgap, then the problem is readily solved. The theory ($S_b$) is then insensitive to
changes in the boundary conditions and $S_{eff}$ still equals $S_e$ except for corrections which
are exponentially small in the system size. Hence, the action $S_e$ really
stands for the {\em fixed point} action of the quantum Hall state. It is the only source of massless excitations
in the problem also in case the $\nu$ is slightly different from being a precise integer. The quantity $m(\nu)$ in $S_e$ 
is now identified as the quantized Hall conductance, at least for small intervals about integral $\nu$.

It is clear that this procedure of generating quantum Hall states cannot be continued indefenitly, to arbitrary
values of $\nu$. At some point the process must stop and {\em transitions} must take place, from one quantum Hall state
to the next. It is important to keep in mind, however, that nothing of this admittedly sketchmatic scenario of
strong coupling physics seems to
depend on the number of field components $N_1$ and $N_2$ in the theory. The basic assumptions that have been used, i.e.
asymptotic freedom and massless edge excitations, are quite generally valid, for all positive values $N_1$ and $N_2$,
including the case of interest $N_1 =N_2 =0$. This, then, is the main motivation of our claim, made in the beginning,
which says that replica field theory can be used as a laboratory for quantum Hall physics.

\subsubsection{Theory of effective parameters}
The action $S_{eff}$, in the limit where $\omega \rightarrow 0$, contains all the information on the
renormalization of the theory and one may proceed in a fashion which is similar to what has been 
outlined several times
before (see Ref. 11). For example, $S_{eff}$ can generally be written as

\be
S_{eff} = S_e [t] +{S'}_b [t]
\ee
where the form of ${S'}_b$ can be obtained from symmetry considerations and the result is

\be
{S'}_b [t] =  -\fr{1}{8} {\qs'}_{xx}\int d^2 x \tr \prt_\mu \tilde{Q} \prt_\mu \tilde{Q} 
 	+\fr{\theta'}{16\pi} \int d^2 x \tr \qe_{\mu\nu} \tilde{Q} \prt_\mu \tilde{Q} \prt_\nu \tilde{Q}
\ee
with

\be
\tilde{Q} = t^{-1} \Lambda t .
\ee
The quantities ${\qs'}_{xx}$ and $\theta'$ are physical observables (in the sense of Gross) and they
may be expressed in terms of correlations of the $Q_0$ fields. Assuming
rotational invariance for simplicity we can write the result as follows

\be
{\qs'}_{xx} = \qs^0_{xx} + \fr{ (\qs^0_{xx} )^2}{8N_1 N_2 L^2} \int d^2 x \int d^2 x'  < tr J_\mu (x) J_\mu (x') >
\ee
and

\be
\theta' =\theta(\nu) + \fr{ (\qs^0_{xx} )^2}{8N_1 N_2 L^2} \int d^2 x \int d^2 x'  < tr \epsilon_{\mu\nu} 
J_\mu (x) J_\nu (x') \Lambda > .
\ee
Here, $J_\mu = Q_0 \prt_\mu Q_0$ equals the Noether current, $L^2$ denotes the area of the system and all 
the expectations  $< ... >$ are with respect to the theory $S_b [Q_0 ]$ with $\omega = 0$.

It is easy to see that the theory of "effective" parameters ${\qs'}_{xx}$ and $\theta'$ and hence ${S'}_b$
generally obey the particle hole symmetry as discussed in the previous Section. In particular,

\be
{\qs'}_{xx} (\nu_+ ) = {\qs'}_{xx} (\nu_- )
\ee
and

\be
\theta' (\nu_+ ) = -\theta' (\nu_- ) .
\ee
These results can be obtained as a direct consequence of the fact that the topological charge of the $Q_0$
matrix field is integer quantized.

\begin{figure}
\begin{center}
\includegraphics[width=8cm]{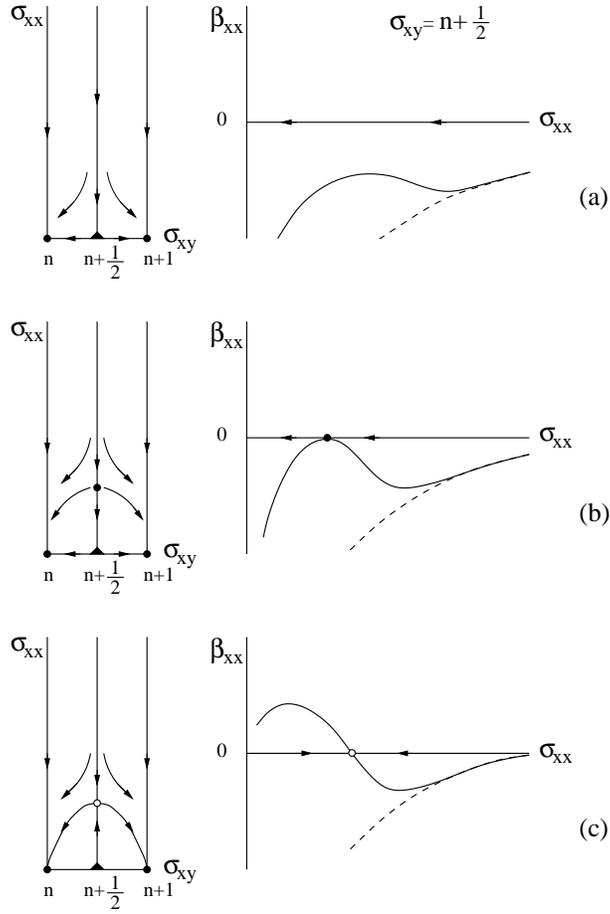}
\end{center}
\caption{
The renormalization group flow diagram for different values of
$N_1$ and $N_2$. (a) The results for large values, (c) the
results
for $N_1 , N_2 \rightarrow 0$. (b) Typical result of the
intermediate
case (see text). The functions $\beta_{xx}$ along the lines
$\sigma_{xy}=
n+\frac{1}{2}$ indicate how the critical fixed point is generated
as $N_1$ and $N_2$ go to small values less than unity.
}
\end{figure}

\subsubsection{Instantons}
The theory of effective parameters has been studied in great detail
before and it can
be used to obtain both the perturbative and non-perturbative (instanton)
contributions to the renormalization.

For later purposes we first give the instanton contribution to the free energy which can be written as

\be
\frac{F_{inst}}{L^2} = D_{N_1 , N_2 }^0 \int \frac{d\lambda}{\lambda^3} (\sigma_{xx}^0 )^{N_1 +N_2} e^{-2 \pi \sigma_{xx}^0
+(N_1 +N_2 ) ln \mu \lambda} cos (\theta (\nu) )
\ee
where the numerical constant $D_{N_1 , N_2}^0$ is given by

\be
D_{N_1 ,N_2 }^0 = \frac{64}{\pi e} \frac{ (\pi / 4\sqrt{e} )^{N_1 +N_2}}{\Gamma(N_1 ) \Gamma(N_2 )}
\ee
Next, recall the complete result for the renormalization group $\beta$ functions which we write as follows

\be
\beta_{xx} =\frac{d\sigma_{xx}}{dlnL} = \beta_{xx}^0 - D_{N_1 , N_2} \sigma_{xx}^{N_1 +N_2 +2} e^{-2\pi\sigma_{xx}} cos \theta ,
\ee

\be
\beta_{\theta} =\frac{d\theta}{dlnL} =  - 2\pi D_{N_1 , N_2} \sigma_{xx}^{N_1 +N_2 +2} e^{-2\pi\sigma_{xx}} sin \theta .
\ee
Here, $D_{N_1 , N_2} = D_{N_1 , N_2}^0 /N_1 N_2$ whereas $\beta_{xx}^0$ stands for the perturbative result 

\be
\beta_{xx}^0 = -\frac{N_1 +N_2}{2\pi} -\frac{N_1 N_2 +1}{2\pi^2 \sigma_{xx}} + ...
\ee

These results describe the exact asymptotics as $\sigma_{xx}$ goes to
infinity. Eq. 27 demonstrates that $\theta$ renormalization is a typical
instanton effect and, hence, it is invisible in ordinary perturbative
expansions.

In order to see the general significance of these results, it is
extremely
important to keep track of the various pieces of strong coupling
information
that we now have. For example, our theory of effective parameters implies
that
the $\beta$ functions can be written, quite generally, as an infinite
series
in discrete topological sectors ($n$) as follows

\be
\beta_{xx} = \sum_{0}^{\infty} d_n cos n \theta
\ee

and

\be
\beta_{\theta} = \sum_{0}^{\infty} c_n sin n \theta
\ee

where the $d_n$ and $c_n$ stand for unknown functions of $\sigma_{xx}$.

These results are a direct consequence of the fact that the topological
charge
of the bulk modes, $q(Q_0 )$, is strictly quantized.  They are no longer
a
consequence of the finite action requirements in semiclassical analyses
alone! Moreover, from our discussion on massless edge modes we know that
the quantum Hall fixed points $\sigma'_{xx} = 0$ and
$\sigma'_{xy} = integer$
generally appear as the only infrared stable fixed points of the theory,
for arbitrary
number of field components $N_1$ and $N_2$.

The functions $d_n$ and $c_n$ typically contribute with instanton factors
$e^{-2\pi n \sigma_{xx}^0}$ such that Eqs 26 and 27 can be used to extract
important general information  on the renormalization of the theory for
finite values of $\sigma_{xx}$.

In Fig. 4 we have plotted the results in terms of a renormalization group
flow diagram
of the conductances $\sigma_{xx}$ and $\sigma_{xy}$. The charateristic
features of
the renormalization  strongly depend on the numerical values of $N_1$ and
$N_2$.

For large values of $N_1$ or $N_2$, the results are hardly any different
from what
one obtains in ordinary perturbative expansions, indicating that the
infrared of the
theory is controlled by fixed points located at  $\sigma_{xx} = 0$ (Fig.
4a).

In the opposite limit, when $N_1$ and $N_2$ go to zero, the instantons
strongly
affect the renormalization of the theory. The results (Fig. 4c) indicate
that the
lines $\sigma_{xy} = half integer$ are the domain of attraction of a
critical
fixed point with a {\em finite} critical value $\sigma_{xx}^*$
which is of order unity.

There is a sharp transition, in the space of $N_1$ and $N_2$, between
these two
different renormalization group scenario's (Fig. 4b). The theory
with $N_1 = N_2 =1$
or, equivalently, the $O(3)$ non-linear sigma model is presumably an
example of
such a boarderline case.

\subsubsection{The theories with large and small values of $N_1 ,N_2$}

A finite value of the critical conductance $\sigma_{xx}^*$ has important
physical
consequences. For example, in the theory of the {\em qHe} ($N_1 = N_2
=0$),
a finite $\sigma_{xx}^*$  ensures that the quantum Hall plateau
transitions behave
in every respect like conventional metal-insulator transitions. Such a
transition is
displayed by the theory in $2+\epsilon$ dimensions. Most of the field
theoretic
apparatus that is used to establish universality, as well as the scaling
behavior of physical
observables, can be applied to the plateau transitions in two dimensions
as well.
The plateau transitions are, in fact, a unique laboratory for
investigating
the more subtle aspects of  metals such as conductance fluctuations,
multifractality
of density fluctuations etc. The main advantage of quantum Hall systems
is the
reduced dimensionality which makes numerical experiments easier.

It is in many ways quite obvious that instanton vacuum, for large values
of $N_1$ and $N_2$,
does not display many of the typical features of disordered metals.
It is nevertheless important to know whether the theory still retains the
fundamental
phenomenon of the {\em qHe}. We have seen that quantum Hall physics is
deeply
rooted into the topology of the problem.
Following the results of the previous Section we now have to deal with
the properties of  strong coupling fixed points and this, clearly, is a
chapter of its own.

Before embarking on the large $N$ expansion we first specify our
instanton results
to the $CP^{N-1}$ theory, which is obtained by putting $N_1 =0$ and $N_2
= N-1$.
Write

\be
\frac{F_{inst}}{L^2} = (\frac{N}{2\pi})^N D_{1,N-1}^0 \int \frac{d\lambda}{\lambda^3} (\frac{2\pi \sigma_{xx}^0}{N} -ln \mu \lambda )^N
e^{-2\pi \sigma_{xx}^0 + Nln \mu \lambda} cos(\theta(\nu)).
\ee

We have added the the quantum corrections to $\sigma_{xx}^0$ in the
first part of the
integrant. This is a well known higher order result in instanton
calculus.

The results at large values of  $N$ are usually expressed in terms of the
rescaled
coupling constant or, equivalently, we can write

\be
\sigma = \sigma_{xx}^0 / N .
\ee
Next, there is the expression for the dynamically generated mass $M$

\be
M= \mu e^{-2\pi\sigma}
\ee
where $\mu$ denotes an arbitrary momentum scale. In the large $N$ limit,
and in terms
of $M$, we can write the instanton as follows

\be
\frac{F_{inst}}{L^2} =\int \frac{d\lambda}{\lambda^3} f_0 f^N cos \theta
.
\ee

Here, the quantities $f_0 = f_0 (\lambda M)$ and $f = f(\lambda M )$ are
generally given as functions of the dimensionless quantity $\lambda M$.
Specifically

\be
f  = -\frac{\sqrt{e}}{8} \lambda M \ln \lambda M;
f_0 = \frac{32\sqrt{2} N^{\frac{3}{2}}}{\pi^{\frac{3}{2}} e}
\ee

Notice that the integrant makes sense only if $\lambda M < 1$. This is
precisely the regime where one expects the semiclassical methodology to
be valid.
However, since instantons enter the theory with an arbitrary scale size
$\lambda$,
one is confronted with the usual infrared problem which, up to date, has not
been truly
resolved. We shall return to the infrared problem at a later stage.

For later purposes, we shall express the $\beta$ functions of the
$CP^{N-1}$
theory with large $N$ as follows

\be
\beta_{\sigma} = \frac{d\sigma}{d\ln L} = \sigma (\beta_0 -
h_\sigma g^N cos \theta )
\ee

\be
\beta_\theta = \frac{d\theta}{d\ln L} = -h_\theta g^N sin \theta .
\ee
Here, the quantities $\beta_0$, $h_\sigma$, $h_\theta$ and $g$ are given
as functions of $\sigma$ as follows

\be
\beta_0 = -\frac{1}{2\pi\sigma} - \frac{1}{2\pi^2 \sigma^2}\ldots  ;
g= \frac{\sqrt{e}}{8} 2\pi \sigma e^{-2\pi\sigma}
\ee
and

\be
h_\sigma = \frac{32\sqrt{2} N^{\frac{5}{2}}}{\pi^{\frac{3}{2}} e} \sigma
;
h_\theta = \frac{64\sqrt{2} N^{\frac{5}{2}}}{\pi^{\frac{1}{2}} e}
\sigma^2 .
\ee

The $\beta_0$ is the only quantity in these expressions that is
perturbative in
both weak coupling and $1/N$ expansions. The instanton contributions,
on the other hand, are non-perturbative both ways.

Since the effects of free instantons vanish exponentially in large $N$,
it is often
assumed, and often quoted in the literature, that the concept of  free
instantons
and instanton  gases is completely irrelevant and misleading
as far as the infrared behavior of  the theory is concerned.
The findings of this paper indicate, however, that exactly the opposite
is true.

We shall show that the large $N$ expansion
provides its own resolution of the aforementioned infrared problem.
Instantons
turn out to be fundamentally important since they facilitate the
cross-over
from weak to strong coupling physics. Moreover, we shall show that the
form of the instanton contributions is actually retained by the theory,
all the way
down to the regime of strong coupling. This gives rise to a novel
understanding of  a dense system of strongly interacting instantons and
it permits us
to interprete the infrared behavior of the sytem in terms of a gas of
nearly
free {\em quasi instantons}.

It is important to stress that, until to date, a proper understanding of
the infrared behavior
of the large $N$ theory has actually been lacking. The historical
analyses merely relied on
heuristic arguments, apparently due to Coleman, which apply to the large
$N$ system in the
thermodynamic limit only. These arguments certainly do not provide a
complete physical understanding of the problem, and they do not say
anything about
the subtleties of finite size scaling and renormalization. Attempts in
this direction have
previously been made, but these have remained largely unsuccessful,
however.

\section{$CP^{N-1}$ model}
\subsection{Introduction}
The $CP^{N-1}$ theory is obtained by taking $Q \epsilon \frac{U(N)}{U(N-1) \times U(1)}$. 
In this case  we can parametrize the $N\times N$ matrix fields $Q$ as follows

\be
Q_{\alpha\beta} = -\delta_{\alpha\beta} + 2 z_\alpha^* z_\beta.
\ee
The $z_\alpha$ stand for $N$ complex scalar fields which obey the constraint
$z_\alpha^* z_\alpha = 1$. In terms of the complex scalar fields $z_\alpha$ we now have
\be
S_b (Q_0 ) = \sigma_{xx}^0 \int \left( \partial_\mu z_\alpha^* \partial_\mu z_\alpha
+ z_\alpha^* \partial_\mu z_\alpha z_\beta^* \partial_\mu z_\beta \right)
-\frac{\theta(\nu)}{2\pi} \int \epsilon_{\mu\nu} \partial_\mu z_\alpha^* \partial_\nu z_\alpha
\ee
which is the $CP^{N-1}$ action in a more conventional notation, with $\sigma_{xx}^0$ replacing the 
usual expression for coupling constant $\frac{N}{g^2}$. 

	The theory in $1+1$ space-time dimensions has been
discussed at many places before. Here we summarize the main steps that lead to
an analysis of the theory at large $N$. For the time being, we ignore the spherical boundary conditions
on the $Q_0$ matrix fields. It is more convenient to follow the historical path and first address 
the problem of infinite
systems, by assuming translational invariance (or Lorentz invariance). This, then, sets the stage
for a detailed discussion on finite size effects, edge currents and topology which will be
the main subject of the remainder of this paper.

\subsection{$U(N)$ invariant saddle point}
By introducing a vector field $A_\mu$ one can write the theory as a $U(1)$
gauge theory as follows

\be
S_b (Q_0 ) = \sigma_{xx}^0 \int d^2 x | D_\mu z_\alpha |^2 + i \frac{\theta (\nu )}{2\pi}
\int d^2 x \epsilon_{\mu\nu} \partial_\mu A_\nu 
\ee
where $D_\mu = \partial_\mu +iA_\mu$ . The original $CP^{N-1}$ action is directly obtained 
from the equations of motion $A_\mu = z_\alpha^* \partial_\mu z_\alpha$. 

\be
q=\frac{1}{2\pi} \int d^2 x \epsilon_{\mu\nu} \partial_\mu A_\nu  .
\ee

We next lift the constraint on the $z_\alpha$ fields by introducing an auxiliarly field 
$\lambda (x)$. Write

\be
S_b \rightarrow S_b + i \sigma_{xx}^0 \int d^2 x \lambda (x) \left( |z_\alpha |^2 -1 \right) 
\ee
then the free $z_\alpha$ fields can be integrated out. We end up with an effective 
theory for the $\lambda$ and $A_\mu$ field variables

\be
S_b (A_\mu , \lambda ) = N \left[ Trln (-D^2 +i\lambda ) -i\sigma \int d^2 x \lambda (x) \right]
+ i \frac{\theta (\nu )}{2\pi} \int d^2 x \epsilon_{\mu\nu} \partial_\mu A_\nu 
\ee
with $\sigma = \sigma_{xx}^0 / N$. Putting $A_\mu =0$, then this effective theory 
has a $U(N)$ invariant stationary point in $\lambda$.
If we replace the $\lambda$ field by

\be
i\lambda (x) \rightarrow M^2
\ee
then the stationary point equation for mass $M$  is obtained as 

\be
\int \frac{d^2 k}{(2\pi)^2} \frac{1}{k^2 + M^2} = \sigma .
\ee
The integral diverges in the ultraviolet. Introducing a momentum cutoff $\mu $ we have

\be
\sigma = \frac{1}{2\pi} ln \frac{\mu}{M}, \\ M=\xi^{-1} = \mu e^{-2\pi\sigma} .
\ee
This shows that for large values of $N$, the theory generates a coupling constant dependent 
massgap $M$ or a finite correlation length $\xi$.

\subsection{$\theta$ dependence at large $N$}
\subsubsection{Introduction}
We proceed by neglecting the fluctuations in the $\lambda$ fields which are of order $N^{-1}$.
The $Trln$ is then expanded as a power series in the $A_\mu$ field. Gauge invariance
requires that the quadratic term is proportional to $F_{\mu\nu}^2$

\be
N  Trln (-D^2 +M^2 ) = N  Trln (-\partial^2 +M^2 )+\frac{N}{48\pi M^2} \int d^2 x F_{\mu\nu}^2 + ... 
\ee
The effective action $S_b$ can therefore be written, to lowest order in the $A_\mu$ fields and 
to leading order in $N$, as

\be
S_b (A_\mu ) = \frac{N}{48\pi M^2} \int d^2 x F_{\mu\nu}^2
+ i \frac{\theta (\nu )}{2\pi} \int d^2 x \epsilon_{\mu\nu} \partial_\mu A_\nu .
\ee

\begin{figure}
\begin{center}
\includegraphics[height=4cm]{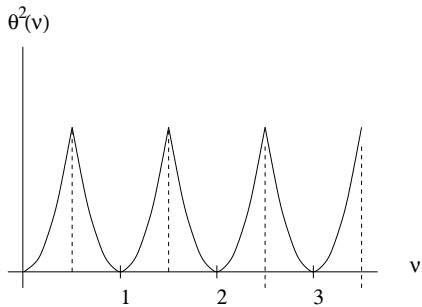}
\end{center}
\caption{The function $\theta^2 (\nu)$ with varying $\nu$ (see text)}
\end{figure}

The $\theta$ dependence of the vacuum free energy is now obtained as  

\be
\frac{F(\theta)}{L^2} =  \frac{3 M^2}{\pi N} \theta^2 (\nu ) .
\ee

In Fig. 5 we have plotted this result as a function of $\nu$. We see that there are 
gusps at half integer values of $\nu$ indicating that the theory undergoes a first
order phase transition. Since the system has a mass, one naively expects that the theory
$S_b$ is insensitive to a change in the boundary conditions (except possibly at
the first order singularities). All that remains of the effective action $S_{eff} (t)$ 
(Section 3.2) is the edge part $S_e$ which now describes the {\em qHe}, 

\be
S_{eff} (t) \rightarrow S_e (t) .
\ee
This means that the $m(\nu)$ in $S_e$ is now apropriately identified as the 
{\em quantized} Hall conductance (see also Fig. 2).

This somewhat cosmetic construction of the {\em qHe} clearly needs a more thorough 
justification and this will be the main topic of the remainder of this paper. 
It is already interesting to notice, however, that the results so far
precisely display Coleman's physical arguments for having periodicity
in $\theta$ (in our notation this corresponds to periodicity in $\nu$).
Recall that in Coleman's picture, $\theta$ describes a background
electric field in $1+1$ space time dimension.
Eq. 45 contains the action of a free electromagnetic field which has been dynamically generated.
It produces a linear potential that confines the $z_\alpha$ particles (bosonic quarks) 
and this explains why the $CP^{N-1}$ model is of interest in particle physics.

It is easy to understand that the linear potential is 
responsible for periodicity in $\theta$. Once $\theta$ passes through odd multiples of $\pi$
it is energetically favorable for the system to materialize charged quark and antiquark pairs from the
vacuum . The quarks and antiquarks
travel to opposite ends of the universe and they set up an electric field such as to 
maximally shield the background electric field $\theta$. 
	 
Notice that the bulk action $S_b (Q_0 )$ contains, by construction, the shielded  
"electric field" $\theta (\nu ) = 2\pi \nu + 2\pi m(\nu)$. Whereas the $2\pi\nu$ stands for the
original or bare value of $\theta$, the $2\pi m(\nu)$ corresponds to the electric field that originates
from the quark antiquark pair in Coleman's language . It is important to remark, however, 
that our theory also contains the dynamics of the "edge" quarks. 
It is described by none other than the familiar massless chiral edgemodes "$t$" in quantum Hall systems.

\subsubsection{Periodicity in $\theta (\nu)$}

After these preliminaries, we next return to the most important issue of this paper, namely 
the quantization of topological charge. Recall that in the original theory that we started out
from, Eq. 41, the topological charge is quantized. In the analysis of the large $N$ theory,
however, no quantization of the topological charge of the $A_{\mu}$ field was assumed. 
This is clearly reflected in the final result 
for the vacuum free energy, Eq. 51,
which is periodic in $\nu$ but not in $\theta (\nu)$ as it should be. 

We shall first proceed in a heuristic fashion and simply enforce a quantized topological charge on
the $A_{\mu}$ field by introducing
another spatially independent field $\lambda'$ according to 

\be
S_b (A_\mu , \lambda' ,n) = \frac{N}{48\pi M^2} \int d^2 x F_{\mu\nu}^2
+ i \frac{\theta (\nu ) + \lambda'}{2\pi} \int d^2 x \epsilon_{\mu\nu} \partial_\mu A_\nu 
+ i\lambda' n.
\ee
Notice that this constraint on the theory does not violate Lorentz invariance.
We next point out that this simple ingredient has several major consequences.
After integration over the free gauge fields we end up with

\be 
S_b (\lambda',n) = - \frac{3 M^2 L^2 }{\pi N} (\theta(\nu) + \lambda')^2 +i\lambda'n .
\ee
Here, $L$ denotes the system size. All spatial dependence has now disappeared from $S_b$. What remains
is the sum over integers $n$ and the integral over $\lambda'$, both of which are elementary. 

\subsubsection{$N \rightarrow \infty$, $L$ fixed}
If one next performs the integral over $\lambda$, then the free energy reduces to a sum over
integers $n$

\be
F(\theta ) = ln \sum_{n} e^{-\frac{\pi N}{12 M^2 L^2} n^2  -i \theta (\nu ) n} .
\ee
The sum is rapidly converging provided $\frac{\pi N}{12 M^2 L^2} >> 1$. This corresponds to the situation
where we let $N$ go to infinity first, keeping $L>>M$ fixed. To leading
order we obtain   

\be
\frac{F(\theta)}{L^2} \approx  \frac{12 M^2}{\pi N} F_0(\frac{\pi N}{12 M^2 L^2} ) cos (\theta (\nu))
\ee
where $F_0(X) = X exp(-X)$. Notice that this type of $\theta$ dependence is a typical instanton result.
By introducing an integral over scale sizes $\lambda$ we
arrive at the same expression as the one found for instantons, i.e.

\be
\frac{F(\theta)}{L^2} = \int_0^L \frac{d\lambda}{\lambda^3} f_0 f^N cos
n\theta
\ee
where the functions $f_0 = f_0 (\lambda M)$ and $f = F (\lambda M)$
are now given by

\be
f_0 = \frac{\pi N}{6 M^2 \lambda^2} - 2
\ee
and
\be
f = e^{-\frac{\pi}{12 M^2 \lambda^2}} .
\ee

The different results that we have obtained for the functions $f_0 $ and
$f$ simply
apply to the different regimes in $\lambda M$ that we have considered. 
When $\lambda M << 1$, these functions are well represented by the
results obtained from instantons. When $1 << \lambda M << \sqrt{\frac{\pi
N}{12}}$
then the expression of this Section apply. Any smooth interpolation
between these
results indicates that the infra red cut off for the instanton gas is set
by
the natural length scale in the problem, i.e. $M^{-1}$.

The most important lesson that one learns from this Section is that by
imposing
a quantized topological charge on the theory, the large $N$ limit and the
thermodynamic limit apparently no longer commute.

\subsubsection{$L \rightarrow \infty$, $N$ fixed}
In order to deal with the limit $L \rightarrow \infty$
we make use of the Poisson summation formula

\be
\sum_{n} e^{i\lambda n} = \sum_{l} \delta (\lambda - 2\pi l)
\ee
with integers $n,l$. The vacuum free energy can now be written as a sum 
over $l$

\be
F(\theta) = -ln \sum_l e^{ -\frac{3 M^2 L^2 }{\pi N} (\theta(\nu) + 2\pi l)^2} .
\ee
This expression rapidly converges for $\frac{3M^2
L^2}{\pi N} >> 1$,
it is obviously periodic in $\theta(\nu)$ and it replaces the result (51)
of the previous Section.

\subsection{$\theta$ renormalization}

We next wish to study the effect of an infinitesimal change $\delta q$ in the integer topological charge,
resulting from a change in the boundary conditions.
This means that we have to evaluate $S_b (\lambda',n+\delta q)$ instead of Eq. 54. Following Section 2, 
however, the edge action now also contributes $S_e (\delta q) = i m(\nu ) \delta q$.
This leads to the following expression for $S_{eff}$, Eq. 16

\be
S_{eff} (\theta, \delta q) = -ln \sum_l e^{ - \frac{3 M^2 L^2 }{\pi N} (\theta(\nu) + 2\pi l)^2 
+2\pi i l \delta q} 
+i m(\nu) \delta q .
\ee
By expanding to linear order in $\delta q$ we identify the Hall conductance as follows
\be
S_{eff} (\theta, \delta q) = F(\theta) +  i{\sigma_{xy}}' \delta q  
\ee
where
\be
{\sigma_{xy}}' = \theta' +m(\nu) .
\ee
In the interval $0<\theta'<\pi$ we have
\be
\theta' = 2\pi \frac{e^{ \frac{12 M^2 L^2 }{ N} \left[ \theta(\nu) - \pi \right]} } 
{1 + e^{ \frac{12 M^2 L^2 }{ N} \left[ \theta(\nu) - \pi \right]}}, 
\ee
whereas for $-\pi < \theta'< 0$ we can write

\be
\theta' =- 2\pi \frac{e^{ -\frac{12 M^2 L^2 }{ N} \left[ \theta(\nu) + \pi \right]} } 
{1 + e^{ -\frac{12 M^2 L^2 }{ N} \left[ \theta(\nu) + \pi \right]}}.
\ee
This, then, is the most important result of this
paper. It shows that the
Hall conductance is quantized.
The corrections to exact
quantization are exponentally small in the {\em area} of the system, 
rather than in the linear dimension. At $\theta(\nu)=\pi$,
however, the large $N$
theory has a gapless phase or, rather, a continuously diverging
correlation length
\be
\xi(\nu) = \xi_0 (\pi - |\theta(\nu)|)^{-\frac{1}{2}}
\ee
where $\xi_0 = \sqrt{\frac{N}{12M^2}}$. The expression for
$\theta'$ is a universal
function of the dimensionless quantity $L/\xi(\nu)$
\be
\theta' = \pm 2\pi \frac{e^{-L^2 /\xi^2 (\nu)}}{1+e^{-L^2/\xi^2
(\nu)}} .
\ee
Here, the + sign holds for $0< \theta(\nu) < \pi$, the $-$  sign for
$-\pi < \theta(\nu) < 0$.

Fig. 6 sketches the results for $\sigma'_{xy}$ and $F(\theta)$ with
varying values of $\nu$.
Notice that  the width of the plateau transition regime $\Delta \nu$ is
inversely proportional
to the area $L^2$ of the system. This means that for geometries like the
infinite strip or
infinite cylinders, the transition is infinitely sharp. The correlation
length $\xi (\nu)$
therefore can not be detected by varying, say, the width of the strip.
The situation is quite different from what one is used to, in
conventional second order
transitions, where $\Delta \nu$ always scales with the shortest distance
in the problem, like the
strip width or the radius of the cylinder. We shall, in what follows,
consider
geometries with a finite area only.

The expression for $\theta'$, or the Hall conductance, can also be
written in differential
form. The result becomes, explicitly written out

\be
\frac{d\theta'}{d\ln L} = \theta'
\bigl[
2- |\frac{\theta'}{\pi}|
\bigr]
{\ln}\frac{|\frac{\theta'}{\pi}|}{2 - |\frac{\theta'}{\pi}|} ,
\ee
This important result demonstrates the meaning of the phrase
$\theta$ {\em  renormalization}. It is no longer dependent on $M$ or $N$
and it's significance is more universal than the
original
expression where it was derived from.
We conclude that $\theta' =0$ is a stable fixed point representing the "quantum Hall effect". 
The $\theta' =\pi$ and $-\pi$ are the unstable fixed points describing the "plateau transitions".
For $\theta$ close to $\pi$ the renormalization group equation can be written as
\be
\frac{d (\pi - \theta')}{d\ln L} =2 (\pi - \theta' ).
\ee
The eigenvalue $2$ equals to the dimension of the system, a well known 
result for first order phase transitions. 

In summary we can say that the large $N$ theory shows all the
fundamental features of the {\em qHe}. Besides
dynamic mass generation and quantization of the Hall conductance,
the theory displays also the massless chiral excitations at the edge
as well as a continuously diverging correlation or localization length
at $\theta = \pi$. We have obtained the exact scaling functions for
the quantum Hall plateau transitions. In the next Section we set up
a more ambitious program that, amongst other things,  enables us to
study the effects of having a finite value of $\sigma'_{xx}$ or,
equivalently,
finite values of the {\em renormalized} coupling constant.

\begin{figure}
\begin{center}
\includegraphics[width=5cm]{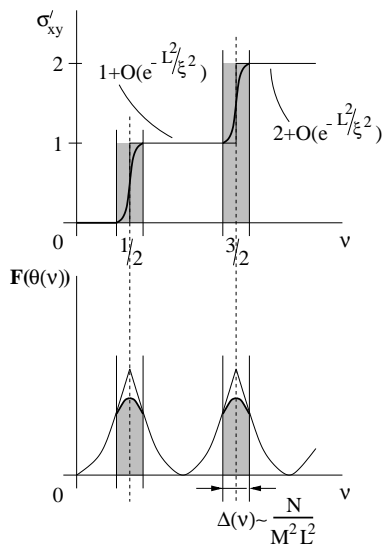}
\end{center}
\caption{The results for the Hall conductance $\sigma'_{xy}$ and
free energy $F(\theta(\nu))$ with varying $\nu$ (see text).}
\end{figure}

\subsubsection{Chiral edge modes}
Notice that $S_{eff}$ really represents the effective action for the "chiral edge modes" $\delta q$.
To be more specific, let us express the change in boundary conditions $Q_0 \rightarrow {t\dagg}Q_0 t$
in $CP^{N-1}$ language. In terms of the vector field $A_\mu$ we have, from the equations of motion

\be
A_\mu = z_\beta^* {t\dagg}_{\beta\alpha} \partial_\mu t_{\alpha\gamma} z_{\gamma}
\ee

Notice that the spherical boundary conditions imply that the $z_\alpha$ reduce to a gauge 
$z_\alpha \rightarrow (e^{i\phi}, 0,0,0,0 ...)$ at spatial infinity. The topological charge $q$
can therefore be expressed as

\be
q = \frac{1}{2\pi} \int d^2 \epsilon_{\mu\nu} (F_{\mu\nu}^0 +f_{\mu\nu})
\ee
where the part with $F_{\mu\nu}^0$ denotes the integral piece of $q$. The second part equals $\delta q$
with 

\be
\epsilon_{\mu\nu} f_{\mu\nu} = \epsilon_{\mu\nu} \partial_\mu a_\nu ; a_\mu =n_\alpha^* \partial_\mu n_\alpha .
\ee
Here, we have introduced the $N$ vector field $n_\alpha = t_{\alpha 1}$ since only the first row of the $t$
contributes. In terms of the $n_\alpha$ effective action for the edge can be expressed as

\be
S_{eff} (n_\alpha ) =  m(\nu) \oint dx n_\alpha^* \partial_x n_\alpha  +h^2 \oint dx |n_1 |^2
\ee
where, as in Section 2, we have added a $h^2$ regulator field.

\section{Coulomb gas}

\subsection{Finite temperatures}

In this Section we set up a computational scheme for finite size scaling. For this purpose we start out by using
the geometry of a cilinder or, equivalently, we may consider a system in $1+1$ space time dimension at finite
temperatures ($\beta^{-1}$). This means that periodic boundary conditions are being imposed
in the imaginary time ($x_0$) direction with period $\beta$. One can next
specialize to the $A_1 =0$ gauge. A straightforeward analysis shows that the  
effective action $S_b$, Eq. 45, now aquires mass terms in the $A_0$ field with a strength of 
order $e^{-\beta M}$. More specifically, the following results are obtained

\be
S_b (A_0 )=  \int_0^\beta dx_0 \int_{-\infty}^{\infty} dx_1 \left[ \frac{N}{48\pi M^2 }  (\partial_1 A_0 )^2
+2 \frac{\sigma_{xx}}{\beta^2} (1 - cos \beta A_0 )  + \frac{i\theta(\nu)}{2\pi} \partial_1 A_0 \right] .
\ee
where
\be
\sigma_{xx} = N \beta \int \frac{dk}{2\pi} e^{-\beta \sqrt{k^2 + M^2}} \approx N \sqrt{ \frac{\beta M}{2\pi} } e^{-\beta M}  .
\ee
Notice that the dimensionless quantity $\sigma_{xx}$ can be interpreted as a measure of the response of the system 
to a change in
boundary conditions in the time direction from, say, periodic to anti periodic. It therefore has the same meaning
as the (dissipative) conductance in Condensed Matter applications. 

\subsection{Finite Universe}

The theory (Eq. 75) is still defined for a strictly infinite spatial extend. 
It has mainly been used as a playground for studying finite action solutions at finite temperatures.
("quantum instantons" or "calorons"). These kinds of semi classical considerations are quite irrelevant, however,
and they do not teach us anything about the behaviour of the system as the temperature goes to zero.

Motivated by our analysis on $\theta$ renormalization we next wish to find a way of handling the mass or
$\sigma_{xx}$ term while introducing, besides a finite temperature $\beta$, also a finite spatial 
extend $L$. For this purpose, we consider an interval $-\frac{L}{2} \le x_1 \le \frac{L}{2}$. In order to define
the theory in the limit $\sigma_{xx} \rightarrow 0$, we fix the $A_0$ field to the set of classical values 
outside the interval, i.e.

\be
x_1 \le -\frac{L}{2} \rightarrow A_0 =\frac{2\pi n_l}{\beta} 
\ee
and

\be
x_1 \ge \frac{L}{2} \rightarrow A_0 =\frac{2\pi n_r}{\beta} .
\ee
Under these circumstances only the interval $-\frac{L}{2} \le x_1 \le \frac{L}{2}$ contributes to the action 
and one can proceed by evaluating the theory in terms of an infinite series in powers of $\sigma_{xx}$.
This theory, which is now defined for a finite {\em area}  $\beta L$ in $1+1$ dimension, still involes infinite sums over 
all possible boundary conditions at $x_1 = \pm \frac{L}{2}$. 

Next, from the detailed analysis that follows one easily concludes that the infrared behavior of the theory is  
generally dominated by those field configurations $A_0$ which are time ($x_0$) independent.
For this reason we consider, from now onward, the $A_0$ field to be independent of $x_0$.  

The topological charge in $S_b$ can be written as

\be
\int_0^\beta dx_0 \int_{-\frac{L}{2}}^{\frac{L}{2}} dx_1 \frac{i\theta(\nu)}{2\pi} \partial_1 A_0 
=\frac{i\theta(\nu)}{2\pi} \left[ \beta A_0^r - \beta A_0^l \right].
\ee
where

\be
A_0^l = A_0 (x_1 = - \frac{L}{2} ) ; A_0^r = A_0 (x_1 = + \frac{L}{2} )
\ee
In order to incorporate the boundary conditions, we can next add lagrange multipliers to this expression. For example,
by adding the following terms 

\be
i\lambda_r \left[ \beta A_0^r -2\pi n_r \right]
\ee
and

\be
i\lambda_l \left[ \beta A_0^l -2\pi n_l \right]
\ee
then the constraints on the $A_0$ field are lifted and the discrete sums over $n_l$ and $n_r$ can be manipulated as
before. By using the Poisson summation formula we obtain the following result for the partition function 

\be
Z=\sum_{m_l ,m_r} Z(m_l ,m_r )
\ee
\be
Z(m_l ,m_r ) = \int [dA_0 ] e^{  -\int dx_1 \left( \kappa (\frac{d\beta A_0}{dx_1})^2 +
2 \frac{\sigma_{xx} }{\beta} (1- cos \beta A_0) +i\beta A_ (x_1 ) \rho_m (x_1 )
\right) }
\ee
where

\be
\rho_m (x_1 ) = (\frac{\theta(\nu)}{2\pi} +m_l ) \delta( x_1 -\frac{L}{2} ) -
(\frac{\theta(\nu)}{2\pi} +m_r ) \delta( x_1 +\frac{L}{2} )
\ee
and

\be
\kappa = \frac{N}{48\pi M^2 \beta}
\ee

\subsection{Classical Coulomb gas}

The remaining steps toward the Coulomb gas representation are quite standard. First, we write the
partition function as an infinite series in powers of $cos\beta A_0$ 

\be
e^{\int dx_1 2 \frac{\sigma_{xx} }{\beta} cos \beta A_0 } = \sum_{n_\pm =0}^\infty 
\frac{ (\frac{\sigma_{xx} }{\beta})^{n_+ + n_-} }{n_+ ! n_- !}
\prod_{i=1}^{n_+} \int dx_i^+ \prod_{j=1}^{n_-}
\int dx_j^-   e^{\int dx_1 i\beta A_0 (x_1 ) \rho_n (x_1 )} 
\ee
where

\be
\rho_n (x_1) = \sum_{i=1}^{n_+} \delta(x_1 -x_i^+ ) - \sum_{j=1}^{n_-} \delta(x_1 - x_j^- ) .
\ee
The integral over $A_0$ can now be performed and the result can be written as

\be
Z(m_l ,m_r ) = \sum_{n_\pm =0}^\infty \delta_{\rho_{mn} ,0}
\frac{ (\frac{\sigma_{xx} }{\beta})^{n_+ + n_-} }{n_+ ! n_- !}
\prod_{i=1}^{n_+} \int dx_i^+ \prod_{j=1}^{n_-} \int dx_j^-
e^{- \frac{1}{8\kappa}  {\cal H}_{coul}}
\ee 
where
\be
{\cal H}_{coul} = -\int dx \int dy \rho_{mn} (x ) |x-y| \rho_{mn} (y )
\ee
represents a system of interacting charges. Here,
\be
\rho_{mn} (x) = \rho_n (x) + \rho_m (x )
\ee
denotes the total charge density whereas $|x-y|$ stands for the Coulomb potential in one dimension. The latter
is obtained in momentum space as follows

\be
|x-y| = -\int dk \frac{e^{ik(x-y)} -1}{k^2}
\ee
where the second term in the integrant is a reflection of charge neutrality of the system. 
Notice that in one dimension the total energy 
of charged excitations diverges linearly in the system size.
The symbol $\delta_{\rho_{mn} ,0}$
indicates that only the configurations with a zero total charge $\int dx \rho_{mn} (x) = 0$ contribute
to the partition function.

The quantity $\frac{\sigma}{\beta}$ is identified as the fugacity of the Coulomb gas. It is straightforward
to verify that the results of the previous Sections are all obtained in the 
Coulomb gas representation by taking the terms with $n_+ =n_- =0$ only, i.e. in the limit of zero fugacity.

\section{Computational results} 

We are now in a position to perform systematic expansions in powers of the fugacity $\frac{\sigma_{xx}}{\beta}$.
For this purpose we generalize our previous notion of "changes in the boundary conditions". We are
interested in evaluating the following partition function 

\be
Z[ a_0^l , a_0^r ] = \sum_{m_l , m_r} Z(m_l , m_r ) e^{i \beta a_0^l m_l   - i \beta a_0^r m_r   } .
\ee
Here, the symbols $a_0^l$ and $a_0^r$ just represent infinitesimal changes in the
boundary conditions on the $A_0$ field. As will be shown below, the results can be written in terms of 
an effective action which 
to lowest order in the $a_0^l$ and $a_0^r$ fields reads as follows

\be
ln Z[ a_0^l , a_0^r ] = F_\theta + \frac{2 L}{\beta} {\sigma'_{xx}} \left( 1 - cos \beta (\frac{a_0^l + a_0^r}{2} ) \right)
+i\frac{{\theta}'}{2\pi} \beta (a_0^l - a_0^r ) .
\ee
Notice that this result is of the same form as $S_b (A_0 )$, Eq. 75, with $\frac{a_0^l + a_0^r}{2}$ now standing for a 
uniform shift in the $A_0$
field and $\beta (a_0^l - a_0^r )$ represents an infinitesimal change in the topological charge.
The quantities ${\sigma'_{xx}}$ and ${\theta}'$ constitute the theory of effective parameters and they have
the same meaning as discussed before (Section 3.2.1). 

\subsection{Effective parameters}
In order to obtain an expression for the quantities ${\sigma'_{xx}}$ and ${\theta}'$
we expand the partition function to lowest orders in powers of the fugacity $\frac{\sigma_{xx}}{\beta}$. By evaluating
the terms in the sum with ($m_l , m_r$)=($m , m$) and ($m \pm 1 , m$) with integer $m$ we obtain 
the following expression
\be
Z[ a_0^l , a_0^r ] = \sum_{m} e^{-\frac{L}{4\kappa} (\frac{\theta(\nu)}{2\pi} + m )^2  +  i m \beta ( a_0^l - a_0^r  )} 
\times \zeta (m)
\ee
where

\be
\zeta (m) = 1 + (4\frac{\kappa}{\beta} \sigma_{xx} ) \frac{e^{i \beta a_0^l} +e^{- i \beta a_0^r }}{\frac{\theta(\nu)}{\pi} +2m -1}
- (4\frac{\kappa}{\beta} \sigma_{xx} ) \frac{e^{- i \beta a_0^l} +e^{ i \beta a_0^r }}{\frac{\theta(\nu)}{\pi} +2m +1} .
\ee
This expression is of the same form as our earlier result which is obtained by taking the $\zeta (m)$ equal to unity. 

Since we are not interested in distinguishing between finite size $L$ and finite temperature $\beta^{-1}$ we shall,
from now onward, put $\beta = L$ for simplicity. As before, we shall consider the cases with large and small values of 
$\frac{L}{\kappa}$ separately. We say that the system is in the {\em quantum Hall phase} when the value of $\frac{L}{\kappa}$ 
is large. The phrase {\em intermediate phase} is being used to indicate systems with a small value of 
$\frac{L}{\kappa}$.

\subsubsection{Quantum Hall phase, $LM >> \sqrt{N}$}
This corresponds to the case where $\frac{L}{\kappa}$ is large. When $0 < \theta (\nu) < \pi$, then only the terms in the sum with
$m=0, -1$ contribute. Write 

\be
{\tilde{\theta}} = \frac{L}{4\kappa} ( \frac{\theta (\nu) }{\pi } -1 ) = (L/\xi)^2 ,
\ee
then the results for ${\sigma_{xx}}'$ and ${\theta}'$ can be written as

\be
\frac{\theta '}{\pi} - 1 = {\tilde{\theta}} F({\tilde{\theta}},\sigma_{xx}) ; 
\sigma_{xx} ' =  \sigma_{xx} F({\tilde{\theta}},\sigma_{xx})
\ee
where
\be
F({\tilde{\theta}},\sigma_{xx}) =  \frac{ \frac{ e^{\tilde{\theta}} -1}{\tilde{\theta}}}
{1+ e^{\tilde{\theta}} + 2 \sigma_{xx} \frac{ e^{\tilde{\theta}} -1}{\tilde{\theta}}} .
\ee
These results can be trivially extended to include the interval $-\pi < \theta (\nu) < 0$. 
Notice that the effective parameters generally depend on two scaling variables, i.e. $L/M$ and
$L/\xi$ respectivily, and they display particle hole symmetry.
Moreover, the renormalization group $\beta$ functions for the quantities
$\sigma' =\sigma_{xx}'/N$ and ${\theta}'$ have the following general form

\be
\frac{d \theta '}{d\ln L} = \beta_\theta (\sigma ' , \theta ' ) ;
\frac{d \sigma'}{d\ln L} = \beta_\sigma (\sigma ' , \theta ' ) .
\ee
We are not quite sure, however, about the consequences of these results for the quantum Hall fixed points.
For example, if we work to lowest order and put $\sigma_{xx} = 0$ in the expression for $F$, then we find,
as before, that the corrections to exact quantization are exponential in the {\em area} of the system. 
However, by working with Eqs 98 and 99 as they stand, with a finite value of $\sigma_{xx}$, we conclude that
$\frac{\theta'}{\pi} \approx \sigma'_{xx} \approx \frac{\sigma_{xx}}{{\tilde{\theta}}} \approx N exp(-LM)$
which is clearly very different. This is possibly a sign for non-perturbative behavior, indicating that the 
infinite orders in the series will have to be resummed.

In what follows, we do not attempt to resolve this problem of the quantum Hall fixed points. In fact, we are going to 
be interested in the overall renormalization behavior of the theory and we keep in mind to stay away from
the singularities at $\theta' = 0$.  

The explicit results for the case where $\sigma \rightarrow 0$ are as follows
\be
\beta_{\theta}=-2\theta'
\bigl[
2-|\frac{\theta'}{\pi}|
\bigr]
{\rm Arth}\big( 1- |\frac{\theta'}{\pi}| \big),
\ee
which we obtained earlier. In addition, we now have also
\be
\beta_\sigma =\sigma'
(\beta_0 + f(\theta'))
\ee
where $\beta_0 \approx \ln \sigma'$ is a function of $\sigma'$ only and 
\be
f(\theta')=2|\theta'|
\biggl[
2-|\frac{\theta'}{\pi}|
\biggr]
\frac{{\rm Arth}(1- |\frac{\theta'}{\pi}|)}{1-|\frac{\theta'}{\pi}|}
+\ln
\frac{{\rm Arth}(1- |\frac{\theta'}{\pi}|)}{1-|\frac{\theta'}{\pi}|} .
\ee

\begin{figure}
\begin{center}
\includegraphics[width=5cm]{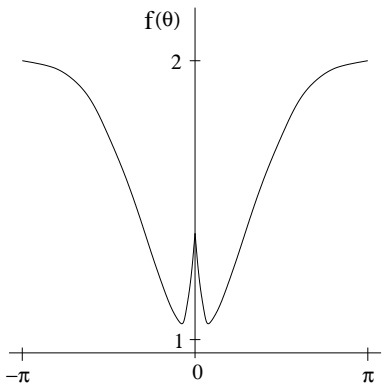}
\end{center}
\caption{The function $f(\theta')$ with varying $\theta'$ (see text).}
\end{figure}\begin{figure}
\begin{center}
\includegraphics[width=5cm]{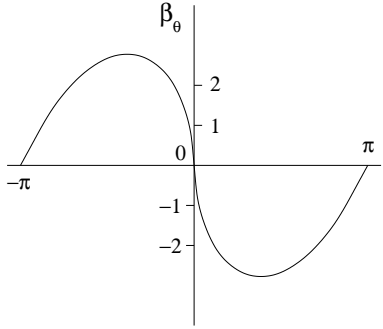}
\end{center}
\caption{The function $\beta_\theta$ with varying $\theta'$ (see
text).}
\end{figure}

The important feature of these results is that they
are independent of $N$.
In Figs 7 and 8 we have plotted the functions
$f(\theta')$ and $\beta_\sigma$ with varying $\theta'$.

   Notice that the corrections to the asymptotic form of the
beta functions generally depend on $N$.
They are obtained by expanding the expressions for
$\theta'$ and $\sigma'$
as a series in powers of the fugacity $\sigma_{xx} = N
\sigma$. For example,
the following result can be obtained for the renormalization of
$\epsilon = \pi - \theta' \approx 0$

\be
\frac{d\epsilon}{d\ln L} = (2-2N\beta_0 \sigma') \epsilon  .
\ee
This represents the lowest order correction to the renormalization group
eigenvalue $2$ as on moves away from the fixed point value
$\sigma'=0$.
The explicit $N$ dependence is just a refection of the fact that in this
problem, the large $N$ limit and the thermodynamic limit do not commute.

It is important to keep in mind, however, that the corrections due to
finite values of $\sigma'$ come from two different sources. Besides
the fugacity $\sigma_{xx}$ there is also the effect of having a finite
value of $L/\kappa \approx \frac{48\pi}{N} {\ln}^2 \sigma$.

In the following Sections we show that the dominant
contributions come from $L/\kappa$.

\subsubsection{Intermediate phase, $\sqrt{N} >> LM >> \ln N$}
In this range of values of $LM$, the $\frac{L}{\kappa}$ can be condisered as a small parameter such that
the Poisson formula can be used in order to obtain a rapidly converging results. At the same time,
the quantity $\sigma_{xx}$ is small such that our expansion in powers of the fugacity is still valid.
The following expressions can be extracted from the Coulomb gas results

\be
\frac{\theta '}{2\pi} = \frac{\theta}{2\pi} - \frac{4\pi \kappa}{L} e^{-\frac{4\pi^2 \kappa}{L}} sin \theta
\ee
\be
\sigma_{xx} ' =  \sigma_{xx} (1 -  e^{-\frac{4\pi^2 \kappa}{L}} cos \theta ) .
\ee
From these expressions one obtains the following renormalization group $\beta$ functions

\be 
\beta_\theta = -h_\theta g^N sin \theta'
\ee
\be
\beta_\sigma = \sigma' (\beta_0 -h_\sigma g^N cos \theta' )
\ee
where $h_\theta , h_\sigma$ and $g$ are functions of $\sigma'$ and given by

\be
h_\theta = \frac{\pi}{6\ln^2 \sigma'} N ; h_\sigma = \frac{ \pi^2}{32 \ln^4 \sigma'} N^2 ; g= exp(-\frac{\pi}{12 \ln^2 \sigma'}) .
\ee
Notice that these expressions are the same as those
obtained from instantons,
but the $beta_0$, $g$, $h_\sigma$ and $h_\theta$ are now
different functions
of $\sigma'$. However, the different results apply to
different regimes in
$LM$ and, hence, to different values of $\sigma'$. The
results of this Section
therefore extend the validity of the instanton methodology to
include both the
weak coupling regime and the intermediate phase of the strong
coupling regime.

\subsection{Discussion}

We have now completed the strong coupling analysis of the $CP^{N-1}$ model with large $N$. 
We have seen that the strong coupling regime ($LM >> 1$) is devided into two different
phases. First, there is the socalled {\em quantum Hall phase} which generally appears at much
larger lengthscales than what one naively would expect, i.e. $LM >> \sqrt{N}$. Secondly,
there is an {\em intermediate phase}, appearing at a set of intermediate lengthscales
$\sqrt{N} >> LM >> \ln N$. In this case the $\theta$ dependence of both the free energy
and the renormalization is quite 
similar to what one obtains in the weak coupling regime, from instantons. 

The results of the large $N$ theory therefore
provide the answer to
the complications which typically arise from instanton calculus but which
can not be handled by the instanton methodology alone.

We next wish to shine some more light on the cross-over from
the intermediate phase
to the quantum Hall phase. In order to discuss this cross-over, one
clearly has to expand the
asymptotic results of Section 6.1.1 as an infinite series in
``topological sectors'' $n$.
We write the results as follows

\be
\beta_\theta = \sum_{n=1}^\infty d_n^0 sin(n\theta')
\ee
and
\be
\beta_\sigma =\sigma'(\beta_0 + \sum_{n=0}^\infty c_n^0
cos(n\theta').
\ee
Here, the coefficients $c_n^0$ and $d_n^0$ are ordinary numbers. They
represent
the $\sigma \rightarrow 0$ limit of the functions $c_n (\sigma')$
and $d_n (\sigma')$
that appear in the general expression of the $\beta$ functions.
For completeness, we have numerically evaluated the lowest order terms in
the series and the results are given by

\be
\begin{array}{lll}
c_0^0=+1.649& \quad& \\
c_1^0=-0.431&      &d_1^0=-2.794 \\
c_2^0=-0.108&      &d_2^0=-0.633 \\
c_3^0=-0.041&      &d_3^0=-0.268 \\
c_4^0=-0.017&      &d_4^0=-0.146 \\
\vdots      &      &\vdots       \\
\end{array}
\ee

The series is rapidly converging in general except at the small region
$\theta' \approx 0$ where the asymptotic $\beta$ functions are
logarithmic singular
(see also Figs 7 and 8). The decomposition in
in discrete sectors $n$ actually works quite well for the unstable fixed
point at $\theta' =\pi$
which  describes the singularities of the quantum Hall plateau
transition.
For example, by keeping only the $n=1$ term in the series for
$\beta_\theta$, we
obtain an approximate value for the renormalization group eigenvalue
which is equal to
2.8 (as compared to the exact value of 2). By truncating the series at a
larger value of $n$
one can get a rapidly converging answer with corrections of the order
$n^{-2}$.   

Since the $n=1$ sector gives the dominant contribution to the
$\beta$ functions,
even in the quantum Hall phase, it is possible to construct a {\em
generalized instanton expression}
for the renormalization that works quite well over the entire regime,
from weak coupling
all the way down to strong coupling. For this purpose, we use the results
of the intermediate phase, Section 6.1.2, and  notice that  the correct
results for the quantum Hall phase are obtained
if we retain the expressions for $\beta_0$ and $g$, but replace
the quantities $h_\theta$ and $h_\sigma$ by the following expressions

\be
h_\theta = -d_0^0 +O(N\beta_0 \sigma')
\ee
and
\be
h_\sigma = -c_0^0 +O(N\beta_0 \sigma')
\ee
Under these circumstances, we obtain the correct asymptotics as
$\sigma'\rightarrow 0$.
Notice that in the same limit, the corrections $O(N\beta_0
\sigma')$ are negligible compared
to the factors $g^N$. 
On the other hand, by retaining the factors $g^N$ in the expressions for
the $\beta$ functions, 
one is ensured that the results always stay finite as $N$ is being sent
to $\infty$.

The generalized instanton expression for the renormalization,
that we have now
completed, displays the most important features of the $CP^{N-1}$ theory
with large
values of $N$. As long as $N$ remains fixed and finite, it gives rise to
a renormalization
group flow diagram in the $\sigma_{xx}$, $\sigma_{xy}$ conductance plane
as sketched in
Fig. 9. It implies that the {\em qHe} is a universal aspect of the system
which  is always
generated, provided the system is large enough.

\begin{figure}
\begin{center}
\includegraphics[height=4cm]{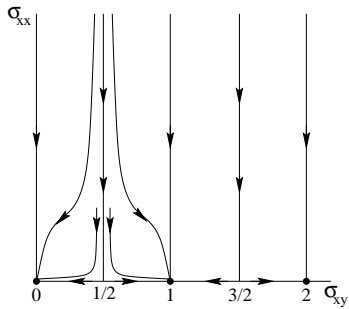}
\end{center}
\caption{The renormalization group flow diagram of the
$CP^{N_1}$ theory with large values of $N$.}
\end{figure}

On the other hand, if we let $N$ go to infinity first, then for all
systems with a finite
value of $\sigma'$, no matter how small, the $\theta$ dependence
disappears and along with
that, the {\em qHe}. This, then, clearly demonstrates the significance of
instantons in the theory
with large $N$.

The procedure can easily be extended to include the
corrections from the topological sectors with $n$ larger than one. The
idea of working with {\em generalized instanton
expressions} for the renormalization group $\beta$ functions impies that
one can interprete
the large $N$ system in terms of  a gas of weakly interacting {\em quasi
instantons} that
have much the same properties as the classical solutions in the weak
coupling phase.
These quasi instantons are not related to the classical ones in any
simple
fashion, however. The large $N$ expansion merely indicates that quasi
instantons
do exist but a more satisfactory answer to the problem would explain this
concept,
starting from an interacting system of topological classical objects.

Finally, we mention the fact that a similar procedure can be followed for
the free energy.
The decomposition in discrete topological sectors $n$ now has a very
different meaning.
In contrast to what happens with the renormalization, the instanton
contributions to the free
energy do not teach us anything about the instabilities of
the theory at $\theta = \pi$. In fact, {\em generalized instanton
expressions} can be
constructed, along  lines that are very similar, but they merely provide
an accurate desciption
of the {\em regular} $\theta$ dependence of the free energy. In this
respect, the instanton
methodology is just an example of  the more general statement which says
that in an
approximate treatment of the problem, free energy considerations alone
are usually not
sufficient. The renormalization, which in our case specifically stands
for {\em $\theta$ renormalization}, generally provides more information
on the singularity structure of the theory  than the free energy itself.

\end{document}